\documentclass[aps, pra, a4paper, amsfonts, amssymb, amsmath, reprint, showkeys, nofootinbib, twoside, superscriptaddress, longbibliography]{revtex4-1}
\usepackage[english]{babel}
\usepackage[utf8]{inputenc}

\usepackage[colorinlistoftodos, color=green!40, prependcaption]{todonotes}
\usepackage{amsthm}
\usepackage{mathtools}
\usepackage{physics}
\usepackage{xcolor}
\usepackage{graphicx}
\usepackage{siunitx}
\usepackage[T1]{fontenc}

\makeatletter
\newcommand{\pushright}[1]{\ifmeasuring@#1\else\omit\hfill$\displaystyle#1$\fi\ignorespaces}
\newcommand{\pushleft}[1]{\ifmeasuring@#1\else\omit$\displaystyle#1$\hfill\fi\ignorespaces}
\makeatother

\newcommand{\eV}{\si{\electronvolt}}
\newcommand{\gcm}{\unit{\gram\per\cubic\centi\metre}}

\usepackage[acronym,nonumberlist]{glossaries}
\newacronym{NP}{NP}{nanoparticle}
\newacronym{GAP}{GAP}{Gaussian approximation potential~\cite{bartok_2010}}
\newacronym{SOAP}{SOAP}{smooth overlap of atomic positions~\cite{bartok_2013}}
\newacronym{fcc}{fcc}{face-centered cubic}
\newacronym{NS}{NS}{nested sampling}
\newacronym{ND}{ND}{neutron diffraction}
\newacronym{EAM}{EAM}{embedded-atom method}
\newacronym{DFT}{DFT}{density-functional theory}
\newacronym{ML}{ML}{machine learning}
\newacronym{MLP}{MLIP}{{\gls{ML}} interatomic potential}
\newacronym{MLIP}{MLIP}{{\gls{ML}} interatomic potential}
\newacronym{PES}{PES}{potential energy surface}
\newacronym{XRD}{XRD}{X-ray diffraction}
\newacronym{HER}{HER}{the hydrogen evolution reaction}
\newacronym{PBE}{PBE}{Perdew-Burke-Ernzerhof~\cite{perdew_1996}}
\newacronym{PBE-DFT}{PBE-DFT}{{\gls{DFT}} with the {\gls{PBE}} exchange-correlation functional}
\newacronym{MD}{MD}{molecular dynamics}
\newacronym{MAD}{MAD}{molecular augmented dynamics}
\newacronym{AIMD}{AIMD}{\textit{ab initio} {\gls{MD}}}
\newacronym{GCMC}{GCMC}{grand-canonical {\gls{MC}}}
\newacronym{HRMC}{HRMC}{hybrid {\gls{RMC}}}
\newacronym{XPS}{XPS}{X-ray photoelectron spectroscopy}
\newacronym{GW}{$GW$}{$GW$ theory}
\newacronym{ANN}{ANN}{artificial neural network}
\newacronym{KRR}{KRR}{kernel ridge regression}
\newacronym{MC}{MC}{Monte Carlo}
\newacronym{GPR}{GPR}{Gaussian process regression}
\newacronym[longplural={core-electron binding energies}]{CEBE}{CEBE}{core-electron binding energy}
\newacronym{ACO}{a-CO$_x$}{oxygen-rich amorphous carbon}
\newacronym{ACH}{a-C:H}{hydrogen-rich amorphous carbon}
\newacronym{ACD}{a-C:D}{deuterium-rich amorphous carbon}
\newacronym{GO}{GO}{graphene oxide}
\newacronym{rGO}{rGO}{reduced graphene oxide}
\newacronym{RMC}{RMC}{reverse {\gls{MC}}}
\newacronym{PDF}{PDF}{pair distribution function}
\newacronym{vdW}{vdW}{van der Waals}
\newacronym{ASE}{ASE}{the Atomic Simulation Environment~\cite{larsen_2017}}
\newacronym{PAW}{PAW}{projector augmented-wave~\cite{bloechl_1994,kresse_1999}}
\newacronym{XANES}{XANES}{X-ray absorption near-edge spectroscopy}
\newacronym{SAXS}{SAXS}{small-angle X-ray scattering}
\newacronym{sp3AC}{$sp^3$ AC}{$sp^3$ amorphous carbon}
\newacronym{taC}{ta-C}{tetrahedral amorphous carbon}
\newacronym{SI}{SI}{supporting information}
\newacronym{PVD}{PVD}{physical vapor deposition}
\newacronym{IR}{IR}{infrared}
\newacronym{NMR}{NMR}{nuclear magnetic resonance}
\newacronym{EDS}{EDS}{experiment-driven simulation}
\newacronym{XAS}{XAS}{X-ray absorption spectroscopy}
\newacronym{RMSE}{RMSE}{root mean square error}

\begin{document}

\title{Molecular augmented dynamics:\\ Generating experimentally consistent atomistic structures by design}

\author{Tigany Zarrouk}
    \email{tigany.zarrouk@aalto.fi}
    \affiliation{Department of Chemistry and Materials Science, Aalto University, 02150 Espoo, Finland}
\author{Miguel A. Caro}
    \email{miguel.caro@aalto.fi, mcaroba@gmail.com}
    \affiliation{Department of Chemistry and Materials Science, Aalto University, 02150 Espoo, Finland}

\date{22 September 2026} % Leave empty to omit a date

\begin{abstract}
A fundamental objective of materials modeling is identifying atomic structures
that align with experimental observables. Conventional approaches for disordered
materials involve sampling from thermodynamic ensembles and hoping for an
experimental match. This process is inefficient and offers no guarantee of
success. We present a method based on modified molecular dynamics, that we call
molecular augmented dynamics (MAD), which identifies structures that
simultaneously match multiple experimental observables and exhibit low energies
as described by a machine learning interatomic potential (MLIP) trained from
\textit{ab-initio} data. We demonstrate its feasibility by finding
representative structures of glassy carbon, nanoporous carbon, ta-C, a-C:D and
a-CO$_x$ that match their respective experimental observables---X-ray
diffraction, neutron diffraction, pair distribution function and X-ray
photoelectron spectroscopy data---using the same initial structure and
underlying MLIP.
The method is general, accepting any experimental observable whose simulated
counterpart can be cast as a function of differentiable atomic descriptors.
This method enables a computational ``microscope'' into
experimental structures.
\end{abstract}

\maketitle

\section{Introduction}

A primary aim of computational materials science is to obtain atomic structures
that explain experimental data. However, many disordered structures generated
from traditional \gls{MD} or \gls{MC} simulations do not agree with experiment:
they sample equilibrium configurations, a condition to which nature is not
bound. Furthermore, there are currently no efficient means by which one can
generate metastable, experimentally feasible structures that are both low in
potential energy and agree with one or more experiments. 

Previous attempts to tackle this problem exist in (hybrid) reverse Monte-Carlo
(RMC/HRMC) approaches~\cite{mcgreevy_1988,opletal_2002}, where Monte-Carlo
trials seek atomic configurations that agree with simple experimental
observables, such as \gls{XRD} or \glspl{PDF}. Similar multi-objective
formulations have been used for crystal structure searches using simulated
annealing~\cite{brunger_1991} and symmetry-constrained genetic algorithm
workflows~\cite{meredig_2013}. However, with all such methods, their domain of applicability has been
restricted to small system sizes of a few thousand atoms with many iterations
being necessary for convergence~\cite{farmahini_2013}. In particular, \gls{HRMC} approaches have historically used simple,
empirical interatomic potentials, that do not closely follow the underlying
\gls{PES}, which further leads to unphysical structures and material-specific bonding
constraints being
necessary~\cite{rigden_1996, gereben_1994, opletal_2013,opletal_2017,khadka_2020,bacilla_2022}.
However, it
has been recently shown that \glspl{MLP} fit to, e.g., \gls{DFT} data, can
circumvent the need for bonding constraints and that more complex observables
can be used with this method~\cite{zarrouk_2024}. 

Despite this success of \gls{MC} approaches, a fundamental problem remains: computational efficiency.
The acceptance rate of \gls{MC}
moves dwindles if atoms find themselves in stable motifs. Hence, simulations
with this method have been restricted to a maximum of a few thousand
atoms~\cite{ishida_2020} and could take over $10^9$ energy evaluations for
convergence~\cite{farmahini_2013}. 
Diffusion models, such as the work of Kwon \textit{et al}.~\cite{kwon_2024}, can predict atomistic structures from input spectra. However, they may not contain enough information about the \gls{PES}, which can lead to unphysical structures even after multiple restarts. Therefore, methods which explicitly include an accurate description of the interatomic potential energy may benefit the search for experimentally feasible structures.

The generalization of inverse structure optimization methods to use \gls{MD}, a far more efficient and
scalable sampling approach than \gls{MC}, therefore seems a desirable step.
However, there have only been a handful of attempts at using \gls{MD} methods to
promote experimental agreement, all of which have
suffered from scaling and/or accuracy issues. For matching \gls{XRD} data,
T{\'o}th and Baranyai proposed reverse molecular dynamics~\cite{toth_2001}.
However, their calculations of partial structure factors, preliminary functions
necessary to calculate the \gls{XRD} diffractogram, used an $\mathcal{O}(N^2)$
scheme without the effects of thermal broadening. Tabulated forces were used due
to this expense, and bond constraints were necessary, which severely hindered
accuracy and applicability. Ishida and Ohba~\cite{ishida_2020} used a
finite-difference method to generate gradients to reproduce the experimental
\glspl{PDF} of carbon blacks. Their calculations produced unstable structures
that were higher in energy than their standard \gls{MD} comparison. Despite
\gls{PDF} calculations typically scaling as $\mathcal{O}(N)$, their experimental
forces were obtained by a central difference scheme, necessitating two
additional \gls{PDF} evaluations for every atom present in the system, which not
only introduced errors in the forces, but also exceptional expense in comparison
to the calculation of analytic gradients. Such analytic gradients were not
available due to the histogram form of the \gls{PDF} which furthermore omitted a
treatment of thermal broadening.
Pitera and Chodera~\cite{pitera_2012} proposed a minimally biased solution to
match experiment through \gls{MC}/\gls{MD} that was also realized through
\gls{EDS} \cite{white_2014, white_2015}. However, its practical application to
large systems and the number of use cases have been limited due to the current
lack of efficient experimental force calculations, available observables, and
its indifference towards time-to-solution. Recent work by Dammann \textit{et
al}.~\cite{dammann_2025} showed liquid-to-solid \gls{EDS} simulations that fit
to \glspl{PDF} using simple forcefields for TiO$_2$ and H$_2$O and the first
formulation of \gls{PDF} gradient forces. However, adherence to minimal biasing
necessitated coupling parameters for each experimental datum, as the biasing
potential \textit{cannot} include the experimental data explicitly: this
information is encoded in the coupling parameters after a large hyperparameter
optimization step specific to each simulation. Furthermore, the simulation time necessary to obtain a viable solution
was on the order of nanoseconds with a non-smooth \gls{PDF} gradient
formulation.  

Common to all \gls{MD} methods that bias to match experiment is that there is a severe lack of scalable, experimental observable prediction models that are amenable to experimental force calculations. The formulations that do exist (only those for \gls{XRD} and \glspl{PDF}) are hindered by inaccurate/non-smooth gradients and a lack of consideration for
thermal effects. Hence, there is a need for reformulation to enable efficient, analytic and
linearly scaling (with respect to the number of atoms) \gls{XRD}/\gls{ND} and
\gls{PDF} predictions which greatly benefit all such methods. Furthermore, a generalization of observable models to those that are descriptor-based---those that are not simple functions of atomic positions---greatly increases the scope of inference that is possible, in particular with \textit{ab-initio} accurate, linearly
scaling potentials. Such work may open up efficient structural inference for some of the most ubiquitous experimental characterization observables. 

In this article, we detail the theory of
\gls{MAD}, an $\mathcal{O}(N)$ modified \gls{MD} method
able to find low-energy, \textit{ab-initio} accurate atomistic structures
that match experimental data. Derivations of the general equations for \gls{MAD}
are given, in addition to $\mathcal{O}(N)$ scaling methods for
\gls{XRD}, \gls{ND}, \gls{PDF} and \gls{XPS} experimental forces which all account
for the important effects of thermal broadening and generate smooth, analytic
gradients. The derivation of \gls{XPS} forces shows that any experimental
observable which depends on local and position-differentiable descriptors is amenable
to this method and is linearly scaling. Concrete scaling tests for CPU and GPU
implementations are shown from the TurboGAP code~\cite{ref_turbogap}.

\section{Theory}
\subsection{Molecular augmented dynamics}

The \gls{MAD} Hamiltonian of an atomic system is given by
\begin{equation}\label{eq:hamiltonian}
\mathcal{H} = T + V + \tilde{V},
\end{equation}
where $T$ is the kinetic energy, $V$ is the interatomic potential and
$\tilde{V}$ is the \textit{experimental potential}.

The dynamics produced by this Hamiltonian result in a multi-objective
optimization of both experimental agreement and the system's (free) energy. The
$\tilde{V}$ term measures some deviation of the current predicted observable(s) from
the experimental data, thereby restraining ``typical'' dynamics to find
configurations that comply with experimental data. The $V$ term hinders the generation
of high-energy atomic configurations that are unphysical.
A system governed by this Hamiltonian can be understood or interpreted as a physical
system (described by $V$) upon which experimental restraints (quantified by $\tilde{V}$)
are imposed or, conversely, as an experimentally compatible system (modeled by $\tilde{V}$) upon which
physical restraints (enforced through $V$) are imposed. In both cases, the kinetic energy
term $T$ controls whether high- or low-entropy configurations are favored during sampling.
We do not explore the connections to thermodynamic ensembles or concepts of statistical physics
in this work, instead introducing the approach phenomenologically on the basis of its usefulness
to build bridges between experiment and simulation. We will try to establish the theoretical
underpinning of the method as a biased-sampling protocol more rigorously in future work.

\gls{MD} with such a Hamiltonian needs
forces from the negative gradients of the potential $V$, which are simply interatomic
forces, and $\tilde{V}$, which are \textit{experimental forces}. To ensure scalability
and computational efficiency, we favor the derivation of analytical forces. We provide the general formalism and specify analytical
gradient expressions for our experimental observables of choice next.

\subsection{General equations for experimental forces}
We wish to induce $M$ experimental observable predictions of an atomic system,
$h^i_{\rm pred}$, to match a corresponding set of experimental data, $h^i_{\rm exp}$.
As such, we define the experimental potential as
\begin{equation}
    \tilde{V} = \frac{\gamma}{2} \sum_{i=1}^M w_i^2\left( h^i_{\rm {pred}} - h^i_{\rm {exp}} \right)^2,
\end{equation}
where $\gamma$ is an energy scale, a factor that determines the importance of
the experimental agreement in comparison to the interatomic potential $V$, and
$w_i$ is a weight which describes the importance/uncertainty associated with
each experimental data point. $\tilde{V}$ simply measures the deviation of
experiment from theory, where we can further define a distance from the predicted observable to the experimental data as
\begin{equation}
D = \sqrt{ \sum\limits_{i=1}^M w_i^2\left( h^i_{\rm {pred}} - h^i_{\rm {exp}} \right)^2 }.
\end{equation}

We can represent this in vector form where elements $[\mathbf{h}_{\text{exp/pred}}]_i = h^i_{\text{exp/pred}}$ are spectrum values associated with a set of discretized dependent variable values and $[\mathbf{w}]_i = w_i$ are the corresponding weights.

Given that the predicted spectrum ultimately depends on the atomic positions, i.e.,
${h}^i_{\rm pred} = {h}^i_{\rm pred}(\left\{ \mathbf{r}\right\})$,
$\tilde{V}$ can be differentiated with respect to the atomic
coordinates $r_k^{\alpha}$, where $k$ is an atom index and $\alpha$ is a
Cartesian component. As such, we can define an \textit{experimental force},
\begin{align}
\label{eq:experimental_forces}
      \tilde{f}_{ k}^{\alpha} &= -\frac{\partial
  \tilde{V}}{\partial r^{\alpha}_k} \nonumber\\ &=   - \gamma \sum_i {w}^2_i \frac{\partial
  {h}^i_{\rm pred}(\left\{ \mathbf{r}\right\})}{\partial r^{\alpha}_k} \left(
  {h}^i_{\rm {pred}}(\left\{ \mathbf{r}\right\}) - {h}^i_{\rm {exp}} \right).
\end{align}

If an experimental observable is a function with compact support,
i.e., it is non-zero within a bounded domain (e.g., a range of energies or $Q$ values), and only relative intensity
is meaningful, then the prediction and the experimental data can be normalized
by the vector norm,
\begin{equation}\label{eq:norm}
    {h}^i =
    \frac{ {\tilde{h}^i}}{
    \sqrt{ \sum\limits_j (\tilde{h}^j)^2 }},
\end{equation}
where ${\tilde{h}^i}$ is an element of the unnormalized spectrum $\tilde{\textbf{h}}$. Then, the gradient of the normalized predicted spectrum is
\begin{align}
\label{eq:forces_compact}
\frac{ \partial {h}^i_{\rm {pred}}
}{\partial r^{\alpha}_k}  =  \frac{1}{|\tilde{\textbf{h}}_{\rm pred}|} \frac{\partial  {\tilde{h}^i}_{\rm pred} }{\partial
r^{\alpha}_k}  - \frac{\tilde{h}^i_{\rm {pred}}}{|\tilde{\textbf{h}}_{\rm pred}|^3} \left( \sum\limits_j
\tilde{h}^j_{\rm pred}  \frac{\partial \tilde{h}^j_{\rm pred}}{\partial
r^{\alpha}_k} \right),
\end{align}
where $|\tilde{\textbf{h}}_{\rm pred}| = \sqrt{ \sum\limits_j(\tilde{h}^j_{\rm pred})^2}$.
Equation~\eqref{eq:forces_compact} can then be used in Eq.~\eqref{eq:experimental_forces}.
Note that the requirement for compact support applies to the portion of the observable that
is to be fitted. Thus, in practice, an experimental observable without compact support can also be
used for \gls{MAD} optimization with this formulation as long as the intensity is arbitrary, the domain of the fit is itself finite and the
observable can be normalized within its bounds, e.g., an \gls{XPS} spectrum where only the
intensities within the interval $[\varepsilon_\text{min}, \varepsilon_\text{max}]$ are considered.

Multiple experimental observables can be used, hence the general equation for the total force on atom $k$, for $L$ sets of experimental observables, is
\begin{equation}
    f_k^{\alpha\text{, tot}} = f^{\alpha}_k + \sum_{i=1}^L \tilde{f}^{\alpha, i}_{k},
\end{equation}
where $f_{k}^{\alpha} = - \partial_{r^{\alpha}_k}V$ are forces from the underlying interatomic potential. The use of multiple observables may aid in reducing the amount of structural degeneracy, particularly in the cases of spectra corresponding to disordered systems.

By obtaining the experimental forces, one can obtain the modified virial stress tensor
$\boldsymbol{\phi}$ and, thereof, the mechanical stress tensor $\boldsymbol{\sigma}$
and thermodynamic pressure $P$:
\begin{align}
\phi^{\alpha\beta, \text{tot}} = & - \frac{1}{2} \sum_{i=1}^N \sum\limits_{j \neq i}^N(r_j^\alpha - r_i^\alpha) f^{\beta,\text{tot}}_{ij},
\nonumber \\
P^\text{tot} = & - \frac{1}{3} \mathrm{Tr}\{\boldsymbol{\sigma^\text{tot}}\} + \frac{N k_\text{B} \Theta}{\Omega},
\nonumber \\
\boldsymbol{\sigma}^\text{tot} = & - \frac{1}{\Omega} \boldsymbol{\phi}^\text{tot},
\qquad \boldsymbol{\phi}, \boldsymbol{\sigma} \in \mathbb{R}^{3\times 3},
\end{align}
where $\Theta$ is the temperature, $k_\text{B}$ is Boltzmann's constant,
$N$ is the number of particles, and $\Omega$ is the volume of the simulation box.\footnote{We avoid using the far more common symbols $T$ and $V$ for temperature and volume to avoid
confusion with the kinetic and potential energies, respectively.}
This enables generalized barostatting with experimental targets. In other words,
\gls{MAD} can optimize the volume and shape of the simulation box to find the ones
compatible with the experimental data.

To demonstrate the validity of these equations, we give explicit expressions for
\gls{XRD}, \gls{ND}, \gls{PDF} and \gls{XPS} below, with the latter showing that any observable
that can be cast as an explicit function of a differentiable descriptor is compatible with this method.
All the mentioned observables have available \gls{MAD} implementations in the TurboGAP code,
with support for multiple (simultaneous) experimental observable optimization.

\subsection{\gls{XRD} experimental forces}
From the theory of Debye~\cite{debye_1915}, the \gls{XRD} intensity $I^{\rm X}$, can be calculated
from a set of interatomic distances, $\{{r_{ij}}\}$, for a particular $Q = 4\pi \sin\theta / \lambda$,
\begin{equation}
    I^{\rm X}(Q) = \sum_i^N \sum_j^N f_i(Q) f_j(Q) \frac{ \sin( Q \, r_{ij}) }{Q \, r_{ij}}, \end{equation}
where $N$ is the number of atoms, $\theta$ is the half-angle of diffraction,
$\lambda$ is the X-ray wavelength, and $f_i(Q)$ is the scattering factor associated
with the species of atom $i$, which can be calculated as a function of $Q$ from analytical functions~\cite{waasmaier_1995}.

The above equation scales as $N^2$, hence it becomes intractable for large system sizes.
However, \gls{XRD} intensities can be calculated in a fashion which
scales as $\mathcal{O}(N)$ and as $n_s(n_s + 1)/2$ with the number of species $n_s$
through the partial \glspl{PDF} of the atomic assembly~\cite{keen_2001,ashcroft_1967}:
\begin{equation}
g_{ab}(r) =  \frac{n_{ab}(r)}{4\pi r^2\, \text{d}r N_a \rho_b },
\end{equation}
where $a,b$ denote species in the system, $n_{ab}(r)$ is the number of atoms of
species $b$ between $r$ and $r + \,\text{d}r$ from species $a$, $N_a$ is the number
of atoms of species $a$ and $\rho_b = c_b \rho$, where $c_b = N_b / N$ and $\rho$
is the number density of the whole system. This sum is only defined up to a given distance $r_{\rm cut}$, which is the cutoff for the neighbors of each atom.
Note that the introduction of a cutoff is the key ingredient enabling $\mathcal{O}(N)$
scaling.

Using the partial \glspl{PDF}, we can calculate the corresponding $n_s(n_s + 1)/2$ partial
structure factors, which are the Fourier transforms of the partial \glspl{PDF}.
Here, we use the Ashcroft-Langreth formalism~\cite{ashcroft_1967},
\begin{align}\label{eq:structure_factor}
S_{ab}(Q)  = &\delta_{ab} + 4\pi \rho (c_{a} c_{b})^{1/2} \nonumber
\\
&\int_0^{r_{\rm cut}} \, \text{d}r \, r^2 \frac{\sin( Q \, r )}{Q \, r} \frac{\sin( \pi \, r / r_{\rm cut} )}{\pi \, r / r_{\rm cut}} [ g_{ab}(r) - 1 ],
\end{align}
where we have introduced a ``window'' function $w(r) = \sin( \pi r / r_{\rm cut} )/(\pi r / r_{\rm cut})$,
which suppresses oscillations in the partial structure factors arising from the introduction of the cutoff $r_{\rm cut}$~\cite{lorch_1969,gutierrez_2002}.

From the partial structure factors, we can obtain the full \gls{XRD} intensity
by a summation over system species types:
\begin{align}\label{eq:xrdintensity}
I^{\rm X}(Q) = &\sum_{a,b}^{n_{\rm s}} f_{a}(Q) f_{b}(Q) (c_{a}c_{b})^{1/2}  \left[ S_{a b}(Q) - \delta_{a b}\right] \nonumber\\ &+ \sum_{a}^{n_{\rm s}} c_{a}f_{a}(Q)^2.
\end{align}

In this form, the predicted \gls{XRD} pattern does not account for thermal broadening which is present in experiment, nor is there an explicit dependence on atomic positions, which is necessary for atomic forces. We can introduce both features into the \gls{XRD} pattern by using a kernel density estimate of the partial \glspl{PDF} with a Gaussian kernel:
\begin{align}\label{eq:kdeg}
\hat{g}_{ab}(r; \{\mathbf{r}\}) = &\frac{1}{\sqrt{2\pi}}\frac{1}{4\pi r^2 N_a \rho_b } \nonumber\\ &\sum_i \sum_{j\neq i}' \frac{1}{\sigma_{ij}} \exp\left( - \frac{ (r - r_{ij})^2}{2\sigma_{ij}^2}  \right) \delta_{s(i),a}\delta_{s(j),b},
\end{align}
where $r_{ij}$ is the distance between atom $i$ and atom $j$, $\sigma_{ij}$ is the width of the Gaussian kernel, which is in general dependent on bond strength~\cite{chung_1997}, and $\delta_{s(k),c}$ are Kronecker deltas which ensure the species of atom $k$, $s(k)$, is equal to that of $c$. The prime symbol in the second summation indicates that the sum is only performed up to the cutoff. In the rest of this article, properties with a circumflex $\hat{}$ are those derived ultimately from this kernel density estimate.

Differentiating Eq.~\eqref{eq:kdeg}, we obtain
\begin{align}\label{eq:pdf_gradients}
\frac{\partial}{\partial r_{k}^{\alpha} } \hat{g}_{ab}(r; \{\mathbf{r}\}) = &\frac{1}{ \sqrt{2\pi}}\frac{1}{4\pi r^2 N_a \rho_b }  \nonumber\\ &\sum_i \sum_{j \neq i }' \Big[ - 2 \delta_{k,i} \left( r^{\alpha}_{j} - r^{\alpha}_{i} \right) \frac{(r - r_{ij} )}{\sigma_{ij}^2} \nonumber\\
&\frac{1}{\sigma_{ij}r_{ij}}\exp\left( - \frac{ (r - r_{ij})^2}{2\sigma_{ij}^2}  \right) \delta_{s(i),a}\delta_{s(j),b}\Big],
\end{align}
from which we can obtain the derivatives of the partial structure factors:
\begin{align}
\frac{\partial}{\partial r_{k}^{\alpha}} \hat{S}_{ab}&(Q; \{\mathbf{r}\}) =  4\pi \rho (c_{a} c_{b})^{1/2} \nonumber\\ &\int^{r_{\rm cut}}_0 \text{d} r  \left( r^2\frac{\sin( Q \, r )}{Q \, r} \frac{\sin( \pi r / r_{\rm cut} )}{\pi r / r_{\rm cut}}  \frac{\partial \hat{g}_{ab}(r; \{\mathbf{r}\})}{\partial r^{\alpha}_k}  \right).
\end{align}
This allows one to obtain the derivative of the \gls{XRD} intensity:
\begin{equation}\label{eq:xrdforces}
\frac{\partial \hat{I}^{\rm X}(Q; \{\mathbf{r}\})}{\partial r^{\alpha}_k} = \sum_{a,b}^{n_{\rm s}} f_{a}(Q) f_{b}(Q) (c_{a}c_{b})^{1/2}  \frac{\partial \hat{S}_{a b}(Q; \{\mathbf{r}\})}{\partial r^{\alpha}_k},
\end{equation}
which furnishes us with atomic forces to allow for fitting to the \gls{XRD} spectrum by substitution into Eq.~\eqref{eq:experimental_forces}:
\begin{equation}
\tilde{f}_k^{\alpha, \mathrm{X}} = - \frac{\partial
  \tilde{V}_{\rm X}}{\partial r^{\alpha}_k}  =  - \gamma \sum_j {w}^2_j  \frac{\partial
  \hat{{I}}^{\rm X}_{\text{j, pred}}}{\partial r^{\alpha}_k}  \left(
  \hat{{I}}^{\rm X}_{\text{j, pred}} - {I}^{\rm X}_{\text{j, exp}} \right).
  \end{equation}

\subsection{Neutron diffraction forces}
We can obtain experimental forces for \gls{ND} by appropriate substitution of the $Q$-dependent X-ray scattering factors in Eqs.~\eqref{eq:xrdintensity} and \eqref{eq:xrdforces} with the mean coherent neutron scattering lengths, $\bar{b}_{a/b}$, and the mean of squared neutron scattering lengths, $\overline{b^2}_{a/b}$. Hence, for a neutron scattering intensity $I^{\rm N}(Q)$ we have
\begin{align}
\hat{I}^{\rm N}(Q; \{\mathbf{r}\}) = &\sum_{a b}^{n_{\rm s}} \bar{b}_{a} \bar{b}_{b} (c_{a}c_{b})^{1/2}  \left[ \hat{S}_{a b}(Q; \{\mathbf{r}\}) - \delta_{a b}\right] \nonumber\\ &+ \sum_{a} c_{a}\overline{b^2}_{a},
\end{align}
and
\begin{equation}
\frac{\partial \hat{I}^{\rm N}(Q; \{\mathbf{r}\})}{\partial r^{\alpha}_k} = \sum_{a b}^{n_{\rm s}} \bar{b}_{a} \bar{b}_{b} (c_{a}c_{b})^{1/2}  \frac{\partial \hat{S}_{a b}(Q; \{\mathbf{r}\})}{\partial r^{\alpha}_k},
\end{equation}
which can then be used with Eq.~\eqref{eq:experimental_forces}.

\subsection{Pair distribution function forces}
One can fit to the total \gls{PDF} extracted from \gls{XRD}/\gls{ND} experiments by modification of the atomically derived pair distribution function by the appropriate scattering factors~\cite{keen_2001}. Here we show the reduced \gls{PDF}:
\begin{equation}\label{eq:pdf_reduced}
    \hat{G}(r; \{\mathbf{r}\}) = \sum_{ab}^{n_{\rm s}}c_a c_b W_{ab} \left[\hat{g}_{ab}(r; \{\mathbf{r}\}) - 1\right],
\end{equation}
with the derivatives given by
\begin{equation}\label{eq:rpdf_gradients}
    \frac{\partial \hat{G}(r; \{\mathbf{r}\})}{\partial r_k^{\alpha}} = \sum_{ab}^{n_{\rm s}} c_a c_b W_{ab}\frac{\partial \hat{g}_{ab}(r; \{\mathbf{r}\})}{\partial r_k^{\alpha}},
\end{equation}
where $W_{ab} = \bar{b}_a \bar{b}_b$ for neutron scattering and $W_{ab} = K_a K_b/\left[\sum_j c_jK_j  \right]^2$ for X-ray scattering~\cite{keen_2001}, where $K_a$ is the effective number of electrons for species $a$. We can use Eqs.~\eqref{eq:pdf_reduced}, \eqref{eq:rpdf_gradients}, \eqref{eq:pdf_gradients} and~\eqref{eq:experimental_forces} to furnish us with the experimental forces. 

\subsection{Local-property (e.g., \gls{XPS}) experimental forces}
We use \gls{XPS} to introduce local-property experimental forces without loss of
generality, as the formulation is virtually identical. \gls{XPS} is a
characterization method which relies on the fact that core-electron binding
energies are primarily dependent on their \textit{local environment}. Different
motifs exhibit different core-electron binding energies. 
An \gls{XPS} model
can be made if this environmental sensitivity of the core electron
can be sufficiently described. Simple analytical models for such a dependence do not exist due to the quantum-mechanical nature of the problem. However, we can use the flexibility of kernel ridge regression to furnish us with a forward model for \gls{XPS} prediction using \textit{ab-initio} reference data~\cite{klawohn_2023}. We use the core-electron binding energy models of Golze \textit{et
al.}~\cite{golze_2022}, which are trained on \gls{DFT} and $GW$ calculations. \gls{XPS} spectrum predictions $\tilde{I}^{\rm XPS} \left( \varepsilon ; \left\{\mathbf{r}\right\}\right)$ are obtained from core-electron binding energies $\varepsilon^i_{\rm pred} =
\varepsilon^i_{\rm pred}\left(\mathbf{q}^i(\left\{ \mathbf{r} \right\})\right)$,
where $i$ is the index of an atom and $\mathbf{q}^i$ is the many-body descriptor of the $i$th atom's local environment (\texttt{soap\_turbo} descriptors~\cite{bartok_2010,bartok_2013,caro_2019}). To account for thermal and experimental broadening, one can smear
out these core-electron binding energies by a Gaussian, resulting in the
following expression for the \gls{XPS} spectrum:
\begin{equation}
     \tilde{I}^{\rm XPS}(\varepsilon; \{\mathbf{r}\})  = \sum_{i} \exp\left(
-\frac{\left(\varepsilon - \varepsilon^{i}_{\rm {pred}}\right)^2}{2\sigma_{\rm XPS}^2}\right).
\end{equation}
Such a model provides an example of an observable which depends on predicted local quantities, in this case $\varepsilon^i_{\rm pred}(\mathbf{q}^i(\{\mathbf{r}\})$, which cannot be expressed as simple functions of atomic positions. 
The critical point to note is that this formulation accepts analytical gradients of the local property. 
Taking the derivative, and exploiting the differentiable nature of atomic descriptors, we obtain:
\begin{align}\label{eq:xps_forces}
 &\frac{\partial \tilde{I}^{\rm XPS}(\varepsilon; \{\mathbf{r}\})}{\partial
r^{\alpha}_k} = \nonumber\\
&\sum_{i} \frac{(\varepsilon - \varepsilon^{i}_{\rm pred})}{\sigma_{\rm XPS}^2}
\frac{\partial \varepsilon^{i}_{\rm pred}}{\partial r^{\alpha}_k}
 \exp\left(
-\frac{(\varepsilon - \varepsilon^{i}_{\rm {pred}})^2}{2\sigma_{\rm XPS}^2}\right),
\end{align}
where,
\begin{equation}
\frac{\partial \varepsilon^{i}_{\rm pred}\left(  \mathbf{q}^i\left( \left\{ \mathbf{r}\right\} \right) \right) }{\partial r^{\alpha}_k} =  \nabla_{\mathbf{q}^i} \varepsilon^{i}_{\rm pred} \cdot \frac{\partial   \mathbf{q}^i\left( \left\{ \mathbf{r}\right\} \right)  }{\partial r^{\alpha}_k}.
\end{equation}
In the case of the model from Ref.~\cite{golze_2022}, $\varepsilon^{i}_{\rm pred}$ is given as a linear
combination of dot-product kernels:
\begin{equation}
\varepsilon^{i}_{\rm pred} (\mathbf{q}^i(\{\mathbf{r}\})) = \varepsilon_0 +  \delta_s^2 \sum_s \alpha_s \left( \mathbf{q}^i \cdot
\mathbf{q}^s \right)^\zeta,
\end{equation}
where $\delta_s$ is an energy scale and $\{ \alpha_s \}$ are the model's fitting coefficients. $\varepsilon_0$ is a constant energy offset that improves the accuracy of the model. $\{ \mathbf{q}^s \}$
are the descriptors of representative atomic environments and $\zeta$ is a positive number (often an integer)~\cite{golze_2022,klawohn_2023}. Thus, the explicit form of the gradient is:
\begin{equation}
\nabla_{\mathbf{q}^i} \varepsilon^{i}_{\rm pred} = \delta_s^2 \zeta \sum_s \alpha_s \left( \mathbf{q}^i \cdot
\mathbf{q}^s \right)^{\zeta-1} \mathbf{q}^s.
\end{equation}
Therefore, any observable which can be predicted via local differential descriptors,
in this case \gls{SOAP} descriptors $\{\mathbf{q}^i\}$, can be used with \gls{MAD}. As \gls{XPS}
spectra exhibit compact support and their absolute intensities are arbitrary,
we can use Eqs.~\eqref{eq:norm},
\eqref{eq:forces_compact} and \eqref{eq:xps_forces} to provide us with \gls{XPS}
experimental forces.
Many other state-of-the-art models for atomic properties/experimental observables take this form~\cite{berger_2024, paruzzo_2018, golze_2022, grisafi_2018, wilkins_2019, grisafi_2021},
making extensions of \gls{MAD} to \gls{XAS}, \gls{NMR}, Raman, and IR spectroscopies
straightforward.

\section{Benchmark methodology}

\subsection{Structure generation}

We propose a series of benchmarks of the \gls{MAD} method through the elucidation of
the atomistic structure of a range of carbon-based materials, both single- and two-species systems.
Carbon materials present an excellent test bed for our methodology for several reasons. Among
these, we highlight the following. Carbon materials are important from a fundamental and
applied perspective. They have been extensively used in the past to test various atomistic modeling
methodologies as they were introduced, including \gls{HRMC} itself~\cite{opletal_2002},
and they possess outstanding structural and chemical complexity and diversity even in chemically pure form.
This means we can focus on testing the capabilities of our method by minimizing extraneous sources
of error like experimental uncertainties, such as the presence of impurities. 
We created experimentally consistent amorphous structures for glassy carbon,
\gls{taC}, \gls{ACD} and \gls{ACO} using \gls{MAD} melt-quench simulations from
randomized structures. The experimental observables used were: \gls{XRD} for
glassy carbon; \gls{ND} for \gls{taC}/\gls{ACD}, and \gls{XPS} for \gls{ACO}. Here, we delve into
the details of the used methodology and detail issues of a technical nature affecting the
accuracy and reliability of the simulations. We also report various scaling results of our algorithms
and CPU/GPU implementations in the TurboGAP code, with which all the calculations were carried out.

The protocol of melt-quench was used as not only is this a standard means for
generating low-energy amorphous structures~\cite{galli_1989,marks_1996,marks_1996b,wang_2022,ibragimova_2025}, such
that meaningful comparisons can be made between simulations with and without
experimental forces, but it also doubled as an optimization method in the case
of \gls{MAD} simulations. This is simulated annealing on an
augmented \gls{PES}. In the following, we refer to the simulations with experimental
forces as ``\gls{MAD}'' simulations and to those without experimental forces simply
as ``control''.

For pure amorphous carbon simulations, we used a carbon \gls{MLP} with \gls{ML}
\gls{vdW} interactions~\cite{muhli_2021} using the Tkatchenko-Scheffler
formalism~\cite{tkatchenko_2009} as the underlying potential. This \gls{MLP} has
been shown to accurately reproduce the \gls{DFT} potential energy surface of
carbon. For \gls{ACD} we used the general-purpose CH \gls{MLP} of Ibragimova
\textit{et al.}
which also includes \gls{vdW} interactions using
Tkatchenko-Scheffler~\cite{ibragimova_2025}. Both used a van der Waals cutoff of $20$~{\AA}. For \gls{ACO} we used the CO
\gls{MLP} from our previous work~\cite{zarrouk_2024} with a \gls{GW}-accurate \gls{SOAP}-based core-electron binding energy model to generate
\gls{XPS} spectra \cite{zarrouk_2024,golze_2022}. Both \glspl{MLP} were validated for accuracy in prior work~\cite{muhli_2021,ibragimova_2025}. 

Each initial configuration was a randomized structure generated by NVT \gls{MD}
simulations using a Bussi thermostat~\cite{bussi_2007}: melting a 27,000-atom cell (10,000 for \gls{ACD})
of diamond at 9000~K for 10~ps, changing the species composition of this cell if
needed (20\% D for \gls{ACD} and 26\% O for \gls{ACO}), and then quenching to 3500~K over 1~ps (5000~K for 1~ps for nanoporous carbon and \gls{ACO}) in the chosen ensemble of the simulation (which could be NVT or NPT). The initial densities for these cells were set to the expected experimental densities: 1.5~{\gcm} for glassy carbon, 3.1~{\gcm} for ta-C, 0.9~{\gcm} for nanoporous carbon, 2.05~{\gcm} for \gls{ACD} and 2.28~{\gcm} for \gls{ACO}. The timestep for \gls{ACD} was 0.1~fs for generating the initial structure and 0.5~fs for other steps. All other simulations used a timestep of 1~fs. All time constants were 100~fs. 
The \gls{MAD} simulations of glassy carbon, ta-C and \gls{ACD} were run under NPT with a Berendsen barostat 
and Bussi thermostat during the \gls{MD} protocol. This allowed \gls{MAD} to find
the volume that matched the experimental data through the calculation of the
generalized virial stress tensor. All other simulations, control or otherwise, were run under NVT conditions with the Bussi thermostat.
Note the choice of a-C:D
instead of the far more common a-C:H, where D stands for isotopically pure
deuterium ($^2$H) and H indicates the hydrogen isotope distribution naturally
occurring in nature, i.e., predominantly protium ($^1$H). The main reason for
this is that quantitatively matching experimentally available ND patterns for
a-C:H is far from trivial due to the large incoherent scattering of protium and
its negative scattering factor in which standard experimental correction
procedures~\cite{placzek_1952,soper_2009} to obtain the coherent portion of the
spectrum break down ~\cite{burke_1993,walters_1994,daykin_2023}. Deuterium has
a similar (positive) neutron scattering factor to carbon and experimental
matching of the ND pattern is more straightforward with a reduction in the
overall error present in the spectrum. This choice still allows us our main
objective for this benchmark: to test the performance of MAD for multicomponent
systems.

\begin{table*}
\begin{center}
\caption{Parameters used for the various \gls{MAD} simulations. a-C:D has species-dependent smearing ($\sigma_{\rm C-C}=0.1$~\AA{}, $\sigma_{\rm C-H}=0.12$~\AA{} and $\sigma_{\rm D-D}=0.14$~\AA{}) to account for the larger Debye-Waller factor of deuterium. Note that the $\gamma$ for the \gls{ACO} \gls{XPS} matching is significantly larger than those for matching diffraction experiments due to the much lower magnitude of the normalized \gls{XPS} spectrum differences.}
\begin{tabular}{l|l|l|l|r|l|l|l}
Material & Observable & Fitting Function & Ensemble & \(\gamma_{\rm max}\) (eV/atom) & \(w(x)\) & \(r_{\rm cut}\) ({\AA}) & \(\sigma\) ({\AA}) \\
\hline
Glassy Carbon & XRD & \(QF(Q)\) & NPT & 0.074 & \(1+\exp(-3.5(x-1.2))\) & 14.1 & 0.1 \\
ta-C & ND &  \(QF(Q)\) & NPT & 0.074 & \(1+\exp(-3.5(x-1.2))\) & 14.1 & 0.1\\
Nanoporous Carbon & PDF &  \(G(r)\) & NVT & 0.037 & \(1\) & 20.0 & 0.1\\
a-C:D & ND & \(QF(Q)\) & NPT & 0.04 & \(1+\exp(-3.5(x-1.2))\) & 14.1 & 0.1--0.14\\
a-CO\(_x\) & XPS & \(\tilde{I}_{\rm XPS}\) & NVT & 7.4 & \(1\) & N/A & N/A\\
\end{tabular}\label{table:parameters}
\end{center}
\end{table*}

Scattering data for glassy carbon, \gls{taC} and \gls{ACD} were extracted from
the papers of Zeng \textit{et al.}~\cite{zeng_2017}, Gilkes \textit{et
al.}~\cite{gilkes_1995} and Burke \textit{et al.}~\cite{burke_1993},
respectively, and were preprocessed, where applicable, with
\textsc{pdfgetX3}~\cite{juhas_2013}. $\sigma_{ij}$ was taken to be a constant despite
it depending, in principle, on the bond strength. The value of $\sigma =
0.1$~{\AA} was used for all scattering simulations for carbon species, and is
close to the values obtained by Gilkes~\cite{gilkes_1995}. More accurate
approximations of this width are possible by applying the Debye-Waller
theorem~\cite{chung_1997,cope_2007}, which in the harmonic approximation would
require an evaluation of the vibrational modes at every timestep. This was not
deemed necessary for the primary focus of the paper which is linear-scaling
experimental forces for augmented dynamics. That said, these widths were made
species-dependent in the case of a-C:D, with $\sigma_{\rm C-C}=0.1$~\AA{},
$\sigma_{\rm C-H}=0.12$~\AA{} and $\sigma_{\rm D-D}=0.14$~\AA{}, where the
latter value was estimated from the work of Daykin~\cite{daykin_2023}. The
sample density for the \gls{PDF} was 21.3~{\AA}$^{-1}$. The experimental
\gls{XPS} data was extracted from the work of Santini \textit{et
al.}~\cite{santini_2015}, and a width of $\sigma_{\rm xps} = 0.4$~{\eV} was
used.
\gls{PDF} data for nanoporous carbon was
taken from Forse \textit{et al.}~\cite{forse_2015}. The cutoff distance for
neighbors was $r_{\rm cut} = 14.1$~{\AA} apart from nanoporous carbon where
$r_{\rm cut} = 20$~{\AA}. 

For \gls{XRD}/\gls{ND}, we fit to $QF(Q)$ to enhance features present at high $Q$, where
\begin{equation}
    QF(Q) = \frac{Q\left[\hat{I}^{\rm X}(Q) - \sum_{a}^{n_{\rm s}} c_{a}f_{a}(Q)^2 \right]}{\left[ \sum_{a}^{n_{\rm s}} c_{a}f_{a}(Q) \right]^2},
\end{equation}
in the case of \gls{XRD}. For \gls{ND}, an appropriate substitution of the $Q$-dependent scattering factors for the mean neutron scattering lengths is necessary.

The following NPT/NVT MD protocol was used for all simulations, with the exception of nanoporous carbon (cf. Fig.~\ref{fig:similarity} a) in Sec.~\ref{section:results}):
1) anneal at 3500~K, while increasing $\gamma$ linearly from 0~{\eV} to $\gamma_{\rm max}$ over 100~ps;
2) cool down from 3500~K to 300~K at $\gamma_{\rm max}$ for 50~ps;
3) anneal at 300~K reducing $\gamma$ linearly from the previous $\gamma_{\rm max}$ value to 0~{\eV} over 50~ps to produce the final structure. Nanoporous carbon and \gls{ACO} structures followed the same timescales as above, but the temperature protocol differed: 5000~K to 3000~K for step 1) and 3000~K to 300~K for step 2). 
For comparison, we performed control simulations with the exact same protocols with the experimental forces turned off and the volume kept at the expected experimental density.

The parameters used in the \gls{MAD} simulations are summarized in
Table~\ref{table:parameters}. 

\subsection{Structural analysis}

Motif analyses were performed by defining bonds in the structure (where a bond
is defined to be within 60~\% of the sum of two \gls{vdW} radii, where $r_{\mathrm{vdW}}^{\rm C} = 1.7$~{\AA},
$r_{\mathrm{vdW}}^{\rm O} = 1.52$~{\AA} and $r_{\mathrm{vdW}}^{\rm D} = 1.2$~{\AA}) and
creating local graphs for each atom in the structure. These local graphs were
composed of the atom of interest and its first (and second, depending on the
motif being sought) nearest neighbors, where each atom was a node and bonds
were edges. A database of graphs corresponding to common motifs (e.g.,
$sp^3$, $sp^2$ and $sp$) was then used to look for subgraph isomorphisms to
determine if a motif was present. These were sought in a hierarchical fashion
from most bonds/species to least bonds/species (e.g., $sp^3 \succ sp^2
\succ sp$). For \gls{ACD} and \gls{ACO}, the subgraph database was extended to
contain common deuterated and oxygenated motifs as in previous
work~\cite{zarrouk_2024,el-machachi_2024}.

\subsection{Scaling}

Scaling tests for the most computationally intensive \gls{MAD} experimental
forces, those of \gls{XRD}, were done on the pre-exascale machine LUMI. CPU results used LUMI-C where each node had 128 cores (2x AMD EPYC 7763
CPUs) with 256~GB RAM (with only 224~GB of RAM actually available) with a peak
theoretical bandwidth of 204.8~Gb/s per socket. GPU results used LUMI-G, where
each node had 64 CPU cores (1x AMD EPYC ``Trento'' CPU) with 4 AMD MI250x GPUs
(which have two compute dies each) with a theoretical bidirectional bandwidth
peak between CPUs and GPUs of 288~Gb/s. Each die has 64~GB of HBM memory
available. The time was taken for the experimental energy and force subroutine
to complete 10 iterations of \gls{MAD} forces in addition to the time for
GPU-to-CPU transfer. These calculations used randomized carbon structures
(generated from diamond and kept at the density of 3.52~{g/cm$^3$}) and were
performed with TurboGAP using the same pure carbon \gls{MLP} as above. All
timings were measured with \textsc{mpi\_wtime}. Strong scaling tests were done
with 27,000 atoms and weak scaling tests were performed from 1,728 to 1,000,000
atoms, both with the number of cores (and GPUs) ranging from 1 to 512. The raw
time results for weak scaling were adjusted by a multiplicative factor given by
the inverse ratio of atoms used by the simulation to that expected by doubling
from 1728 atoms (all structures used were actually significantly greater in size than that of
raw doubling from 1728 atoms due to the structure generation method from repeating diamond unit cells).
The implementation is that of OpenMPI for CPU and OpenMPI+HIP+HOP for GPU, where HOP is
a header-only porting library to allow interoperability between CUDA and HIP
architectures. Batching of computations was implemented for the GPU code, with
the number of batches being provided by the user, to reduce memory usage for
large systems with limited resources (e.g., a single GPU). All GPU kernels are
explicit with the only dependency being that of cuBLAS/hipBLAS for their respective
architectures. 

\subsection{Gradient implementation}

\begin{figure}
    \centering
    \includegraphics[width=\linewidth]{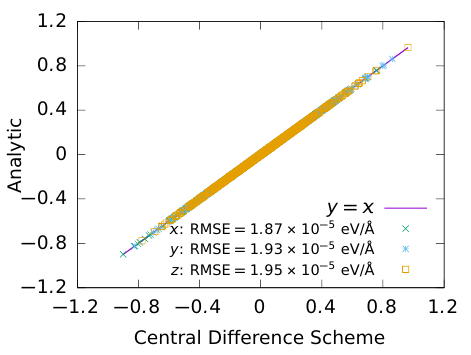}    
    \caption{Analytical forces for \gls{XRD} of a 747-atom CO system (512 C, 235 O) compared to a central difference scheme, in eV/\AA{}. The displacement for the central difference calculations was $\pm 0.001$~{\AA}. 
    }
    \label{fig:central_difference_scheme}
\end{figure}

Finite differences using a central difference scheme were used to validate the analytical gradients for the experimental forces, see Fig.~\ref{fig:central_difference_scheme}. The TurboGAP master branch was used (commit reference 9027b3e)~\cite{ref_turbogap}. As seen from the negligible \glspl{RMSE} reported in the inset, the analytic implementation of the modified forces, incorporating the experimental term, follows the exact gradient of the modified \gls{PES} to within numerical precision.

\section{Results}\label{section:results}

\begin{figure*} 
\begin{minipage}[]{0.7\linewidth}
    \includegraphics[width=\linewidth]{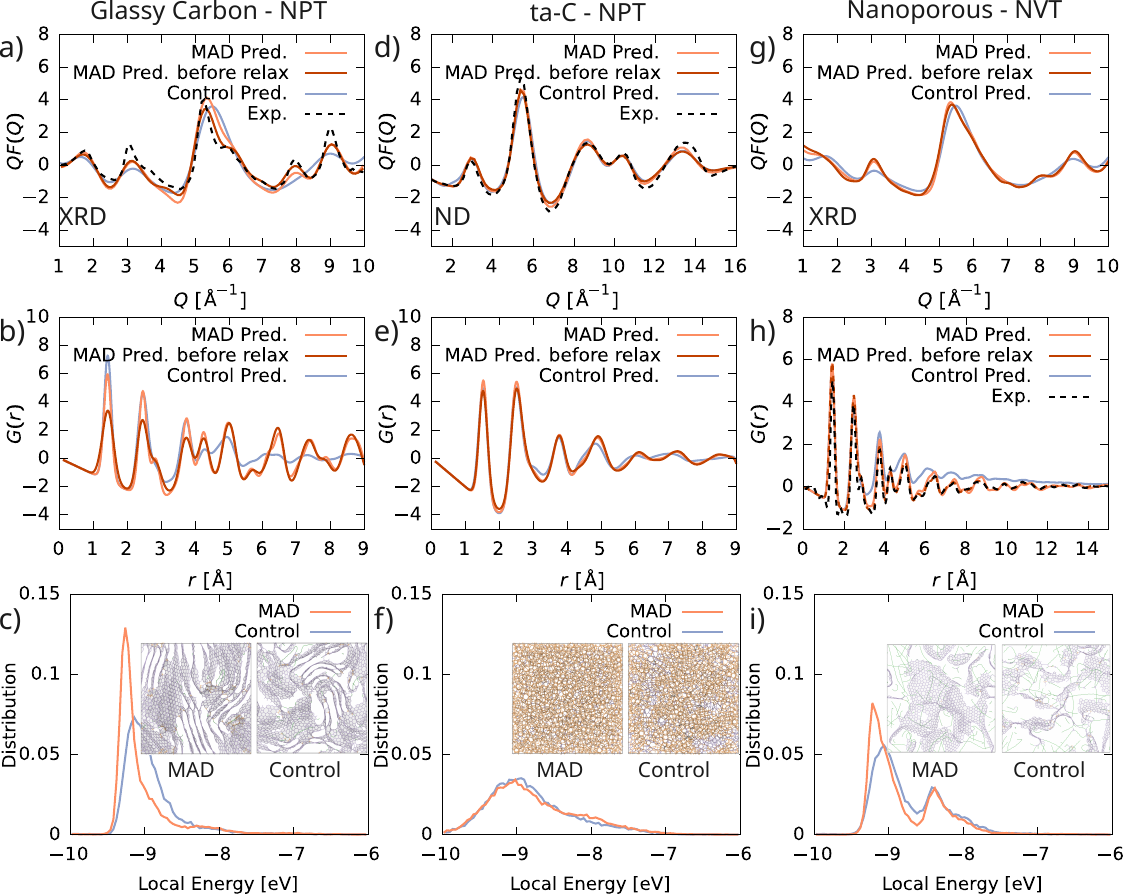}%
\end{minipage}
\hfill
\begin{minipage}[]{0.27\linewidth}
    \caption{XRD/ND patterns, reduced pair-distribution functions,
    structural slices and local-energy distributions of MAD simulations
    compared to control MD simulations. The left column corresponds
    to glassy carbon XRD fitting, the central column corresponds to ta-C
    ND fitting and the right column corresponds to nanoporous carbon
    reduced PDF fitting. a), d) and g) are predictions of scattering data. b),
    e) and h) are the reduced PDFs. c), f) and i) are the MLIP local-energy
    distributions, with 10~{\AA} wide structural slices of the MAD
    simulation compared to the control simulation. $sp^3$, $sp^2$ and $sp$ motifs are colored in
    orange, purple and green, respectively. The
    local-energy distributions show that MAD simulations generate stable
    structures, which are lower or similar in energy to MD
    simulations.}
     \label{fig:xrdglassycarbon}
\end{minipage}
\end{figure*}

\begin{figure}
    \centering
    \includegraphics[width=1.0\linewidth]{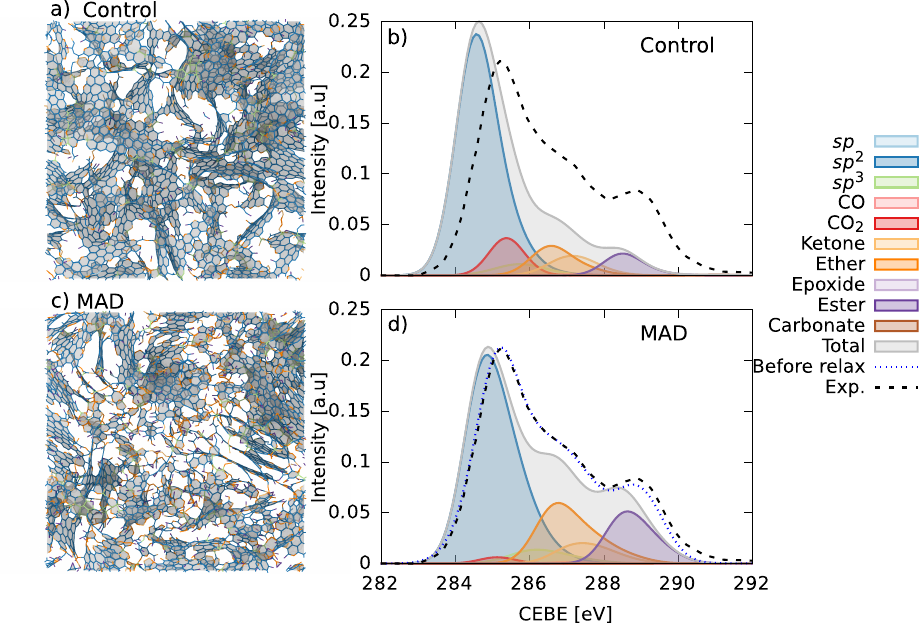}
    \caption{XPS spectra of control and \gls{MAD} simulations of \gls{ACO}
    compared to experimental data \cite{santini_2015}. a), c) 15~{\AA} wide
    slices, with CO$_2$ and O species removed for clarity, of control and MAD
    final structures, respectively. Colors represent different functional groups
    which correspond to the legend on the right. b) Control simulations produced
    a lot of CO$_2$ from the melt-quench procedure, giving structures which do
    not match the XPS spectrum. d) MAD-predicted XPS spectra agrees well with
    experiment by disincentivizing CO$_2$ production. The final spectrum deviates
    slightly from the almost perfect agreement of the spectrum prior to final
    relaxation without the influence of experimental forces.}
    \label{fig:xpsoptim}
\end{figure}

\Acrlong{MAD} produced a range of disordered carbon-based structures that agreed
with the experimental data used, and with experimental
carbon motif percentages and densities, see Fig.~\ref{fig:xrdglassycarbon}, and Table~\ref{table:motifratios}. Structures found with \gls{MAD} were lower in interatomic potential energy than
those of the control simulations, apart from ta-C and \gls{ACO}, where the
energy was marginally higher, but there was better experimental agreement.
This is consistent with the out-of-equilibrium synthesis process in these materials,
which proceeds via high-energy \gls{PVD}~\cite{robertson_2002,santini_2015}.
The ability of \gls{MAD} to produce lower energy final structures
highlights the potential usefulness of \gls{MAD} for overcoming kinetic barriers that may
be inherent to regular \gls{MD}-based sampling protocols.

Densities for NPT \gls{MAD} simulations were close to the experimental values.
For a-C:D, the experimental density was not given, hence the average density of
a-C:H of 30\% hydrogen was used (with the hydrogen mass replaced by that of
deuterium) for the initial structures and control simulations. The
experimental density for the \gls{MAD} simulation was optimized by fitting to
the experimental data. Thus, we can use the \gls{MAD} simulation as a means to infer
the experimental density.

For glassy carbon, curved graphitic (so-called ``turbostratic'') sheets compose
the bulk of the structure, with $sp^3$ motifs linking the sheets together, and
fullerene-like motifs, which agrees with experimentally derived models of glassy
carbon \cite{harris_2004,harris_2005,uskokovic_2021}. \gls{MAD} found a higher
density, 1.9~{\gcm}, compared to that of the control, 1.5~{\gcm}. \gls{MAD}
glassy carbon structures reproduced all the peaks of the \gls{XRD} pattern,
including the high-$Q$ peaks which were not resolved by the control simulations.
This resulted in longer-ranged correlations in the \gls{PDF}, suggesting
significant inter-layer structure correlation. 

The derived \gls{taC} structure is almost entirely composed of $sp^3$ motifs, with
$91.3\%$ $sp^3$ motifs for the \gls{MAD} structure compared to $71.3\%$ for the
control, the former of which agrees well with the $\sim84\%$ proposed by the
authors of the original experiment with a simple coordination analysis
based on the \gls{PDF}~\cite{gilkes_1995}.
The \gls{MAD} and control \gls{ND} spectra are
similar, but with a better reproduction from \gls{MAD}, particularly for low
$Q$, which determines the structure at longer length scales. This results in a
slightly higher density, 3.2~{\gcm}, in comparison to the experimental density,
3.1~{\gcm}.

Both the control and \gls{MAD} melt-quench simulations for nanoporous carbon
resulted in too-high amounts of $sp$ carbon motifs. However, due to the
inclusion of experimental data in the optimization of the structure, the
\gls{MAD} simulations sought lower energy structures, with lower energy $sp^2$
motifs being found and a more connected global structure. It is worth noting
that, experimentally, highly energetic motifs like $sp$ carbon tend to become
passivated either by long-term annealing or reactivity with O- and H-containing
atmospheric species. We have not aimed for capturing this level of chemical
complexity for this study, as the focus here is on presenting the method and
showcasing diverse examples. That said, detailed \gls{MAD} simulations would be
an obvious choice for simultaneously elucidating complex structure \textit{and}
chemistry in disordered materials. Precisely, the next two examples are designed
for evaluating the ability of \gls{MAD} to handle wider chemical spaces by
introducing (separately) oxygen and deuterium into the carbon simulations.

For \gls{ACO}, \gls{MAD} simulations were able to produce the metastable
distribution of motifs which are present in experimental \gls{XPS} data (Fig.~\ref{fig:xpsoptim}). By
contrast, standard control melt-quench simulations resulted in large amounts of
CO$_2$ formation which stymied the reproduction of the secondary peak present in
the \gls{XPS} data. \gls{MAD} inhibited the formation of the thermodynamically
stable CO$_2$ in favor of the formation of carboxyl and ether groups, which
dominate the high core-electron binding energy region of the spectrum in this
material, and are compatible with the out-of-equilibrium physical-deposition
procedure used in the experiment~\cite{santini_2015}. These results echo our
previous work using hybrid reverse Monte Carlo to reproduce the \gls{XPS}
spectrum of \gls{ACO}~\cite{zarrouk_2024}. Notably, with \gls{MAD} we are able to
study significantly larger simulation boxes and obtain the results with lower
computational cost, thanks to the computational advantage of \gls{MD} vs
\gls{MC} ($\mathcal{O}(N)$ vs $\mathcal{O}(N^2)$).

\begin{figure}
    \centering
    \includegraphics[width=\linewidth]{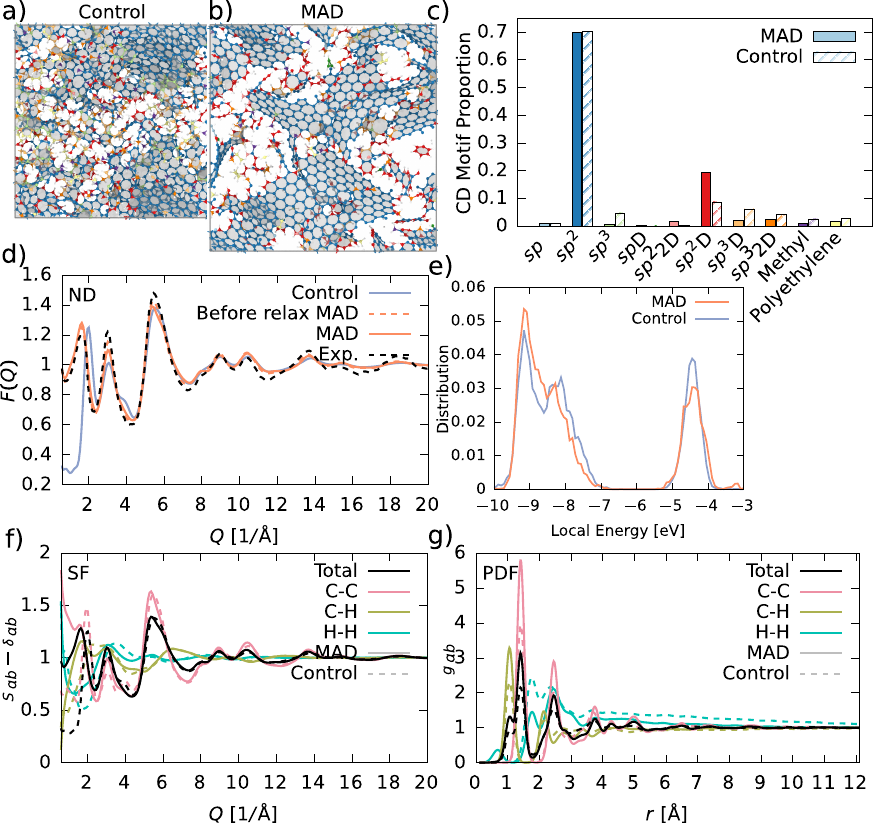}
    \caption{\gls{ACD} \gls{MAD} and control comparison. a) and b): structures
    for control and \gls{MAD} simulations, with atoms colored by motif type,
    following the scheme in c), the proportion of carbon motifs in structures.
    d) \gls{ND} comparison between \gls{MAD} and control. e) Local-energy
    distribution. f) Partial structure factors (without the influence of
    scattering factors) on the \gls{ND} diffractogram in d). g) \glspl{PDF} of
    structures showing increased first and second shell correlations.  }
    \label{fig:acdx}
\end{figure}

In our final example,
\gls{MAD} simulations produced lower energy \gls{ACD} structures that almost
perfectly reproduced the \gls{ND} spectrum of experiment, which is in contrast
to the control simulations, which deviated strongly in the low-$Q$ region
($Q<2$~{\AA}$^{-1}$), see Fig.~\ref{fig:acdx}. The experimental density
was not given for \gls{ACD}, and so the mean density of
a-C:H~\cite{casiraghi_2005} for compositions of $\sim 30\%$ H was used for the
control NVT simulation, whereas the simulation with \gls{MAD} was performed
under NPT to find the density. This demonstrates the power of
\gls{MAD}: the experimental density can be found from the number density
information implicitly contained in the experimental scattering data due
to the construction of the
virial tensor from the experimental forces. Many hydrogenated amorphous carbons
that have $\sim 30\%$ H content are denoted as diamond-like a-C:H, with a high
C-C $sp^3$ content. However, we found, with this annealing time, that
high C-C $sp^2$ coordinated structures were indeed compatible with the \gls{ND}
spectrum of Burke \textit{et al.}~\cite{burke_1993}, suggesting that such structures are
commensurate with the model of graphite-like a-C:H. 

\begin{center}
\begin{table*}
\caption{Table of motif percentages and densities from simulations. Control values are in brackets. Note that the percentages for \gls{ACD} do not account for all the motifs in the structure. These can be seen in our accompanying Letter~\cite{zarrouk_2025}. }
\begin{tabular}{l|l|l|l|l}
Structure & \(sp^3\) (\%) & \(sp^2\) (\%) & \(sp\) (\%) & Density (g/cm\(^3\))\\
\hline
Glassy Carbon & 2.8 (1.5) & 95.2  (92.5) & 2.0 (5.9) & 1.9 (1.5)\\
ta-C & 91.3 (73.4) & 8.7 (25.8) & 0.0 (0.0) & 3.2 (3.1)\\
Nanoporous & 0.2 (1.5) & 70.5 (69.1) & 29.2 (29.4) & 0.9 (0.9)\\
a-C:D & 0.6 (4.7) & 69.9 (70.1) & 1.0 (0.7) & 1.42 (2.05)\\
\end{tabular}
    \label{table:motifratios}
\end{table*}
\end{center}

\begin{figure}
  \centering
  \includegraphics[width=\linewidth]{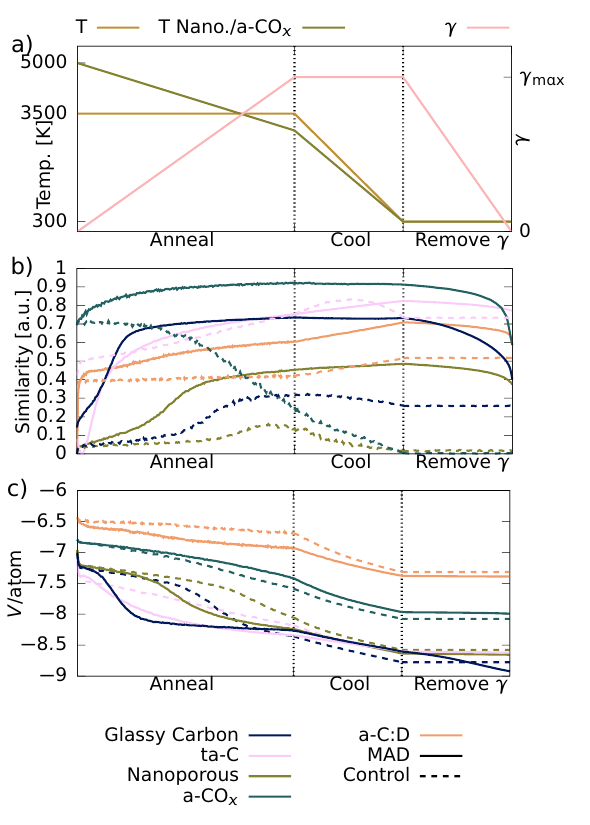}
  \caption{ a) MAD simulation protocol. b) Experimental similarity, $1- D/D_{\rm max}$, from the experimental potential. c) \gls{MLP} potential energy per atom.   }
  \label{fig:similarity}
\end{figure}

\gls{MAD} can accelerate the annealing process to find low-energy structures.
The protocol for both control and \gls{MAD} simulations is depicted
in Fig.~\ref{fig:similarity} a), along with the evolution of the experimental
similarity, defined as  $1 - D/D_{\rm max}$, where $D_{\rm max} =
\mathrm{max}(D_{\rm max}^{\rm control}, D_{\rm max}^{\rm MAD})$ for each
material, Fig.~\ref{fig:similarity} b), with the interatomic potential energy
for \gls{MAD} simulations in Fig.~\ref{fig:similarity} c). For all structures
other than \gls{ACO}, the interatomic potential energy of \gls{MAD} simulations
during annealing decreases much faster than the control simulations, while
exhibiting a concurrent increase in experimental similarity. The similarity gradually increases during the cooling stage while the interatomic potential
energy decreases. Upon gradual removal of the experimental potential, the
interatomic potential energy and similarity decrease in tandem as the system
relaxes. The similarity of control simulations varies depending on the material
and the stage of simulation. However, all final similarities are lower than
\gls{MAD} simulations, and all final potential energies of \gls{MAD} structures,
other than those of \gls{ACO} and ta-C, are lower than the control simulations. 

\begin{figure}
  \centering
  \includegraphics[width=\linewidth]{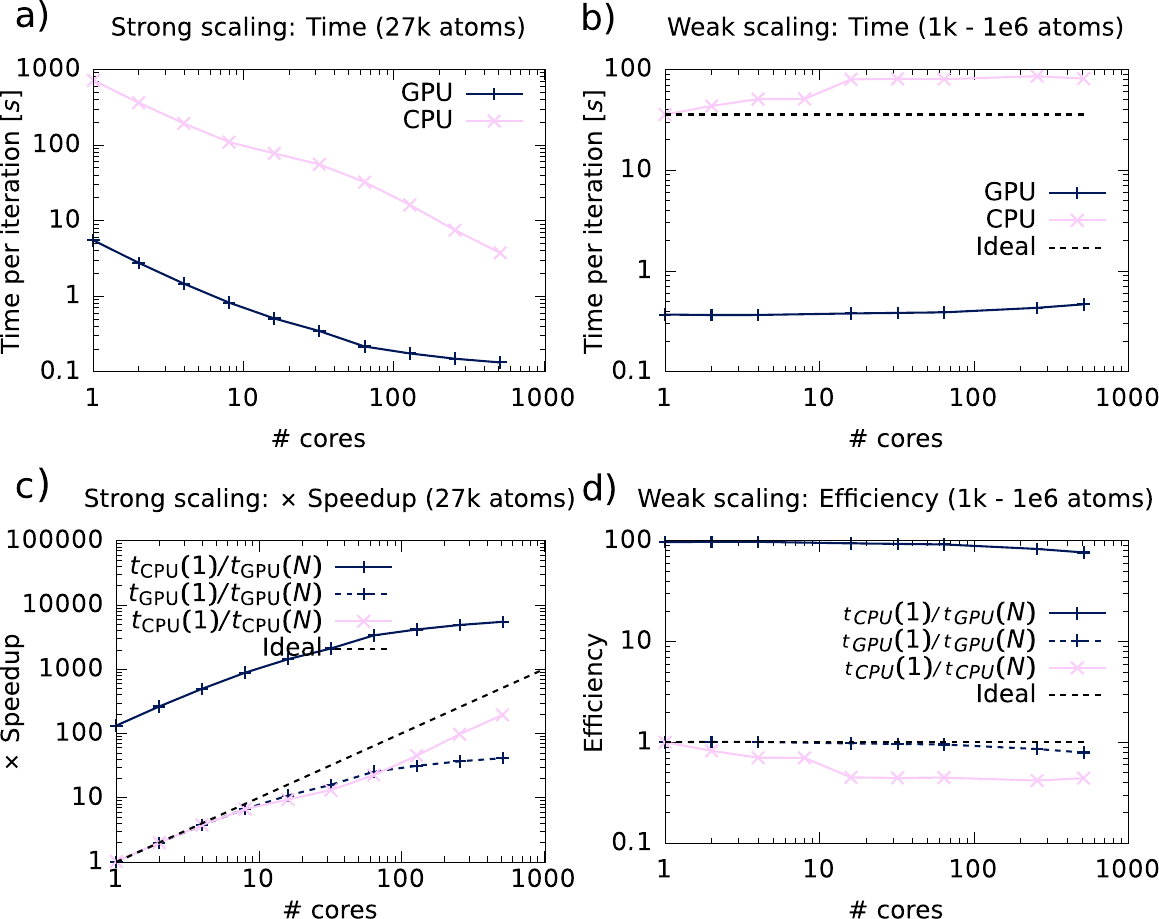}
  \caption{Scaling tests for the most computationally expensive \gls{MAD}
  forces: \gls{XRD}. a,b) Raw times for the subroutines to calculate the
  experimental energies and forces for strong scaling (keeping the number of
  atoms constant but doubling the number of cores and GPUs, if applicable) and
  weak scaling (doubling the system size and doubling the cores and GPUs),
  respectively. c,d) strong scaling speedup and weak scaling efficiency.
  Ideal program results are with respect to the CPU times.}
  \label{fig:scaling}
\end{figure}

The scaling of \gls{MAD} is linear for both CPU and CPU+GPU implementations, with
a significant performance gain from the GPU implementation, as can be seen in
Fig.~\ref{fig:scaling}. Such performance increases render the calculation of
experimental forces marginal in comparison to other CPU heavy tasks in
the code. Deviations of the strong scaling from that of ideal program execution were
primarily due to overheads from MPI broadcasting and GPU to CPU
communication. The weak-scaling performance of the GPU implementation far
outstrips the CPU in terms of efficiency. The efficiency is primarily due to the
calculation of the \gls{PDF}, which is ``embarassingly parallel'', in addition to
cu/hipBLAS operations for the explicit Fourier transform at specific $Q$
points. 

\section{Discussion}

In this work, we have focused on introducing the \gls{MAD} methodology and testing it on
challenging, interesting and diverse disordered carbon-based materials. While relatively simple,
in terms of the number of chemical species involved, these materials have complex chemistry
because of the variety of chemical bonds that carbon atoms can form among themselves and
with oxygen. Therefore, we expect that \gls{MAD} should perform similarly well for other
material systems, and our group is currently working on applications to phase-change materials
and metal oxides, which will be reported elsewhere.

\subsection{Sampling/optimization protocols}

The simulation protocol that we have utilized for these showcase benchmarks was that of
a simple annealing process, which is standard
in melt-quench simulations and thus facilitated comparison with existing, established methodologies.
However, this could be enhanced by using multiple
annealing steps, such as by combining NPT and NVT steps in a more sophisticated fashion.
In principle, there is no limitation of the introduced \gls{MAD} methodology versus established \gls{MD}
protocols, as the modified Hamiltonian formalism with analytical gradients and virial stress tensor
ensures that all the typical simulations that can be performed as part of \gls{MD} workflows can also
be performed within \gls{MAD} with experimental restraints.
For global optimization, using this simulated annealing approach would likely be more efficient if combined with \gls{HRMC} moves or by using \gls{MD}-based methods such as minima hopping~\cite{goedecker_2004}. The greater efficiency of \gls{MAD}, being able to find reasonable structures in the simulated annealing, may be understood by disconnectivity graphs of the biased energy surface in comparison to the standard \gls{PES} of the \gls{MLP}~\cite{de_2019}.

\subsection{Availability and quality of experimental data}

In practice, the most obvious limitation is the availability of reliable experimental observables. Here,
the main bottleneck is the preprocessing of experimental data so that a one-to-one comparison with the
simulated counterpart, needed in the formulation of the modified Hamiltonian, can be made. This involves
removing or accounting for experimental noise, thermal effects and other instrumental or methodological
artifacts inherent to the experimental technique in question. Another limitation is the availability
of accurate and efficient ways to simulate these experimental observables, including linear-scaling behavior
with the number of atoms in the system. In this paper, we have presented relevant methodological frameworks
for ubiquitous characterization methods, namely diffraction and X-ray photoelectron spectroscopy, the latter formulation
being amenable to other (scalar) local observables. The difficulty to extend \gls{MAD} support to other
methods will greatly depend on the nature of the physical interactions taking place, but it can benefit from
existing computational prediction approaches. Prime examples include vibrational (\gls{IR}, Raman) spectroscopies,
\gls{NMR}, molecular probes (e.g., for pore measurements), other X-ray spectroscopic techniques,
and even microscopy. Our group is currently undertaking some of these tasks.

\subsection{Computational prediction of observables}

Another major limitation is the availability of data-driven approaches to predict some of the computed
quantities. In particular, accurate and fast \glspl{MLP} are needed that cover the relevant configuration
space. While ready-to-use \glspl{MLP} may not always be available for the system of choice, the emerging
paradigm of ``foundation'' models~\cite{batatia_2025, mazitov_2025b, yang_2024, lysogorskiy_2026} followed by ``fine-tuning'' offers hope that almost \gls{DFT}-accurate
\glspl{MLP} will be either readily available or easy to make in a semi- or fully automated way. While
some observable predictions benefit from tabulated per-species parameters (like \gls{XRD} and \gls{ND}
scattering factors), a similar issue as for \glspl{MLP} remains for certain observables (like \gls{XPS}):
a specialized \gls{ML} model is needed that requires human time and expertise, and (often expensive) data
to train. Again, the emerging foundation model paradigm offers hope but with more reservations given the
more specialized nature of these models, which limits the scope of community-driven efforts from which
foundation \gls{MLP} models have benefited~\cite{jain_2013,dunn_2020}.

\subsection{Effect of enforcing MAD constraints}

Turning our attention to the technical details of the \gls{MAD} methodology, we are in its infancy with regard to learning ways to tune the various parameters and optimize the simulations, e.g., to obtain a good balance
between the imposed experimental restraints and the physical model.
For instance, as we introduce more potential energy into the system with the
experimental potential, we promote/prevent the formation of certain bonds, depending on
whether they are favored by the experimental potential, the physical potential energy (modeled by the
\gls{MLP}), or both. Gently increasing $\gamma$ to $\gamma_{\rm max}$ allows for reasonable
configurations of bonds before the influence of the experimental potential
enhances sampling of other regions of configuration space. During the annealing process, the
action of the experimental potential becomes stronger, preventing certain bonding configurations. 

\begin{figure}
  \centering
  \includegraphics[width=\linewidth]{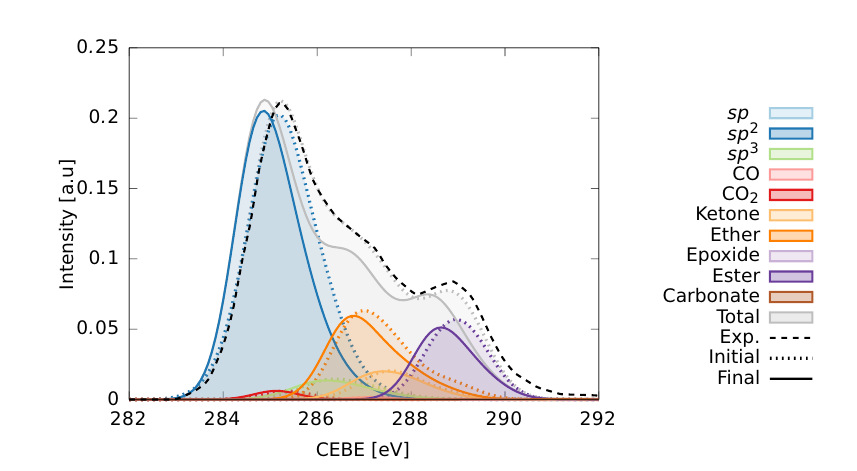}
  \caption{ Shift in \gls{XPS} spectrum from the initial configuration prior to the removal of the experimental potential, to the final configuration. The shifts are generated by \gls{MAD} forces which displace oxygen slightly to be in more local environments which raise the core electron binding energies of many motif types. Such displacements are small and do not significantly affect the interatomic potential energy. }
  \label{fig:xpsshifts}
\end{figure}

\begin{table}
\begin{center}
    \caption{ Table of motif re-classifications for \gls{ACO} from the initial
    configuration prior to removing the experimental potential to the final
    configuration, where the experimental potential has been fully removed.
    Classifications are based on the bond-based, subgraph isomorphism method. We
    see a significant amount of ester to ketone and ether to $sp^2$ transitions,
    suggesting that \gls{MAD} induced small displacements of motifs in bound,
    oxygen-rich sites, to be close to either $sp^2$ or ketone motifs (thus
    making them ethers or esters). The oxygen in these environments raised their
    core electron binding energies, to produce a shift to align with the
    experimental data. Upon removal of the experimental potential, the oxygens
    reconfigured to be in their final stable state (ketones or $sp^2$ sites).
    These results are in line with
    the results of previous work~\cite{zarrouk_2024}. }
\begin{tabular}{lll|r}
\multicolumn{3}{l}{Motif re-classifications} & Count (\%)\\
\hline
$sp^1$    &  $\rightarrow$  &  $sp^2$ & 0.03\\
$sp^2$    &  $\rightarrow$  &  $sp^1$ & 0.01\\
$sp^2$    &  $\rightarrow$  &  $sp^3$ & 0.04\\
$sp^3$    &  $\rightarrow$  &  $sp^2$ & 0.03\\
CO$_2$    &  $\rightarrow$  &  CO & < 0.01\\
Ketone    &  $\rightarrow$  &  Ether & 0.01\\
Ketone    &  $\rightarrow$  &  Ester & 0.01\\
Ether     &  $\rightarrow$  &  $sp^1$ & 0.01\\
Ether     &  $\rightarrow$  &  $sp^2$ & 0.3 \\
Ether     &  $\rightarrow$  &  Ketone & 0.18 \\
Ether     &  $\rightarrow$  &  Epoxide & < 0.01\\
Ether     &  $\rightarrow$  &  Ester & < 0.01\\
Epoxide   &  $\rightarrow$  &  Ether & 0.01\\
Ester     &  $\rightarrow$  &  CO$_2$ & < 0.01\\
Ester     &  $\rightarrow$  &  Ketone & 0.23\\
Ester     &  $\rightarrow$  &  Ether & 0.06\\
Ester     &  $\rightarrow$  &  Carbonate & < 0.01\\
Carbonate &  $\rightarrow$  &  Ester & < 0.01\\
\end{tabular}\label{table:reclassification}
\end{center}
\end{table}

\subsection{System-specific considerations}

Another relevant observation is present in the \gls{ACO} \gls{XPS} \gls{MAD} results:
the similarity drops to its
minimal value at the end of the removal of the experimental potential. This is
due to a shifting of the spectrum from almost perfectly matching the
experimental result, before relaxation of $\gamma_{\text{max}}$, to its final state, in which the
morphology of the spectrum is retained yet core-electron binding energies are shifted down, as seen in Fig.~\ref{fig:xpsshifts}, and the interatomic potential energy decreases slowly upon the removal of the experimental potential. We can
explain this in the context of our previous work~\cite{zarrouk_2024}: oxygen
increases the core-electron binding energies of carbon when it is present in its
local atomic environment. Thus, the experimental data has a shift upwards
in core-electron binding energy due to latent oxygen in the environment of carbon. This
shift led to a potentially erroneous interpretation of the motif structure inside the
material in the original work~\cite{santini_2015}. Due to this shift in the data, \gls{MAD} forces
induced small displacements which allowed for more oxygen to be within the
local environments of carbon. The signatures of this can be seen in the
re-classification of carbon sites in the deconvolution, given in
Table~\ref{table:reclassification}. Overall, less than 1\% of all motifs from the initial configuration are reclassified in the final one after the final structure relaxation, hence the small change in interatomic potential energy. Larger oxygen contents may allow for better experimental agreement. 

Careful consideration of errors present in the processing of experimental data
is paramount in using \gls{MAD}. Such evaluation was necessary in choosing a
deuterated, rather than hydrogenated, amorphous carbon spectrum for \gls{ND}
matching. In diffraction experiments, the structurally relevant quantity to be
extracted is the elastic coherent scattering. This gives the static structure
factor of the material $S(Q)$. However, the raw data obtained in experiment is the
double differential cross section $\frac{\text{d}^2\sigma}{\text{d} \Omega \, \text{d} \omega}$:
the number of neutron counts scattered into solid angle $\text{d}\Omega$ with energy
$\hslash (\omega +\text{d}\omega)$. This has inelastic contributions---where the
material can gain/lose energy---that must be subtracted. For heavy elements, as
the mass of the neutron is far less than that of the element, the energy that a
neutron can transfer/gain from material interaction is small, and therefore the
inelastic corrections of Placzek~\cite{placzek_1952} can be applied to obtain the coherent diffractogram. However, as
protium and deuterium have masses comparable to that of a neutron, the energy
transfers can be as large as the energy of the neutron itself. As such,
inelastic effects induce a $Q$-dependent incoherent background to the
measured differential cross-section~\cite{soper_2009} and the aforementioned
corrections cannot be made. Experimentalists therefore resort to
approximate polynomial correction schemes which can introduce systematic errors.
Protium ($^1$H) has a very large incoherent neutron scattering cross section
($\sim 80$~barn) whereas deuterium ($^2$H) has a significantly smaller incoherent neutron
scattering cross-section ($\sim 2$~barn). As such, the systematic errors present in
the final coherent scattering pattern from deuterated materials are far less than
those of simple hydrogen. 

The need for nuanced analysis of the results for some material systems and experimental
techniques is unlikely to be removed altogether in the near future. Therefore, while we
expect \gls{MAD} to lower the barrier for adoption of an integrated computational+experimental
approach for materials research, there will remain a requirement for domain expertise and,
especially, familiarity with the limitations of the experimental techniques used. Particular attention must be paid to whether the observable is rapidly changing during acquisition, in which case an ensemble-averaged force would be necessary~\cite{pitera_2012,white_2014}. If the observable is inherently non-local, then forward predictive models may be rather costly due to the lack of linear scaling, which would inhibit experimental force calculations at every timestep. 

\subsection{Crystal-structure search}

\gls{MAD}, in its current form, may not be well-suited for crystalline systems. 
Amorphous systems have many energetically degenerate metastable states, which is further
exacerbated by system size. Crystalline structures are exponentially localized
in the free energy landscape due to the small configuration-space volume
associated with their order. In molecular dynamics, high-entropy products are
favored due to the action of temperature; therefore, the sampling of such small
volumes is unlikely during sampling with annealing. Methods that enforce certain symmetries, such as the FPASS
method~\cite{meredig_2013}, would be beneficial for small-enough unit cells in
such cases. A synthesis of the FPASS and MAD method could potentially yield
more efficient results than either method alone, due to the
acceleration provided by the gradients in \gls{MAD}. In this context, one of the obvious advantages of the \gls{MAD} formulation is that analytical, linear-scaling gradients enable flexible reuse of the rich body of existing structure-optimization and structure-sampling methods.

\subsection{Software}

In terms of computational implementation and software availability,
the significant performance of the GPU code renders the calculation of \gls{MAD}
diffractogram energies and forces to be negligible in comparison to the CPU
code, and also lower than the time execution of the \gls{MLP} calculation.
The time for such calculations on the CPU are comparable, or more
expensive, to the calculation of \gls{SOAP} descriptors used in our \glspl{MLP}, which is the most
computationally demanding portion of a typical TurboGAP calculation. Although both CPU and GPU
implementations of \gls{MAD} have been developed in the context of developing the TurboGAP code,
there is no methodological limitation in terms of using \gls{MAD} with other simulation
engines; in fact, we have plans to interface TurboGAP's \gls{MAD} routines with popular \gls{MD}
simulation engines in the near future, thus reaching a significantly wider potential user base.

In conclusion, we believe that the objective of fully integrating experiments and atomistic modeling
is one step closer with the introduction of \gls{MAD}. This is part of the \gls{ML}-driven
advances of recent years that are revolutionizing computational materials science. We expect that
this field will continue to develop at an increasing pace over the next few years.

\section{Outlook}

As we have shown, introduction of the experimental potential into the modified
Hamiltonian enables sampling configurations that agree more favorably with
experimental data. As the experimental data used should correspond to sound,
low-energy metastable configurations, the multi-objective optimization by
\gls{MAD} finds lower interatomic potential energy structures than traditional
\gls{MD} within the same simulation timescale for the pure carbon structures,
with a more agreeable distribution of structural motifs.
The ability of \gls{MAD} to find metastable structures is demonstrated in particular
for \gls{ACO}. The modified potential energy surface inhibited the formation of
thermodynamically favorable products (CO and CO$_2$)
and promoted the formation of more oxygenated motifs. 
Yet, \gls{MAD} simulations do not perfectly match the experimental data, and
the level of agreement may depend on the material and experimental observable at hand.
This is expected for multiple reasons, as follows. 

First, there could be finite-size effects present in the simulation, despite their
current scale. The structures generated by \gls{MAD} are representative
structures which aim to contain the correct statistical distribution of
structural motifs in the material that is implicit in the experimental data.
However, due to the much larger scale of the real material in comparison to the
simulations, and the finite time of optimization, there may be
some distributions of motifs that are not sampled by the optimization process. 

Second, there are errors/noise associated with the measurement of experimental
data which are not accounted for.
This is exacerbated by the idiosyncrasies among measurement devices: one device
may give different systematic errors in comparison to another. In this work, the
experimental data was assumed to be perfect, with no assumption of errors due to
the lack of such information. However, this can be corrected either by generating
weights which reflect the error associated with a particular data point,
e.g., $w_i \propto 1/\sigma_i^{\rm err}$ where $\sigma_i^{\rm err}$ is the
expected error of that particular data point, or by the use of an additional model that
can add device-specific noise to the predicted spectra (and by extension the
experimental potential). The inclusion of this error correction was not included
in this work despite the central role of the experimental data in these models. 

In the case of scattering data, there may further be errors in the Fourier
transform from the \gls{PDF} to the \gls{XRD}/\gls{ND}. Such errors could be
accounted for by another corrective model which can apply a mapping from the
predictions of well-known, ordered structures, to that of the experiment (with
the addition of a model which accounts for the experimental noise).
For \gls{XPS}, the model is currently that of a Gaussian kernel density estimate
from the predicted core-electron binding energies. In general, it should be a
double convolution, with Gaussian and Lorentzian functions which account for
instrumental and thermal broadening, respectively. 

As famously put by Richard Feynman, ``if it disagrees with experiment, it is
wrong''~\cite{feynman_1964}. With this work, we have developed a tool
that can help elucidate the vast multitude of structures that do agree, hoping to
bring computational materials science one step closer to the elusive
experimental reality.

\begin{acknowledgments}
The authors acknowledge financial support from the Research Council of Finland
under projects 330488, 347252, 352484, 355301 and 364778, from the European
Union's M-ERA.NET 3 program (NACAB project under grant agreement No 958174). We
acknowledge the EuroHPC Joint Undertaking for awarding this project (XCALE
innovation study within the Inno4scale project under grant agreement No
101118139) access to the EuroHPC supercomputer LUMI, hosted by CSC (Finland) and
the LUMI consortium through a EuroHPC Regular Access call. The authors also
acknowledge other computational resources from CSC -- the Finnish IT Center for
Science and Aalto University's Science-IT project.
\end{acknowledgments}

\def\bibsection{}
\section*{References}

\end{document}